\documentclass[aps,prd,twocolumn,superscriptaddress,nofootinbib,floatfix,eqsecnum]{revtex4}

\pdfoutput=1

\usepackage{amsfonts}
\usepackage{amsmath}
\usepackage{amssymb}
\usepackage{graphicx,color}
\usepackage{float}
\usepackage{hyperref}
\usepackage{subfigure}
\usepackage{dcolumn}

\begin{document}

\title{Modified gravity versus shear viscosity: imprints on the
  scalar matter perturbations}

\author{C. M. S. Barbosa}
\affiliation{Universidade Federal do Esp\'irito Santo, N\'ucleo Cosmo-ufes \& Departamento de F\'{\i}sica,
  Av. Fernando Ferrari, Goiabeiras, 
29075-910, Vit\'oria, ES, Brasil}

\author{H. Velten}
\affiliation{Universidade Federal do Esp\'irito Santo, N\'ucleo Cosmo-ufes \& Departamento de F\'{\i}sica,
  Av. Fernando Ferrari, Goiabeiras, 
29075-910, Vit\'oria, ES, Brasil}

\author{J. C. Fabris}
\affiliation{Universidade Federal do Esp\'irito Santo, N\'ucleo Cosmo-ufes \& Departamento de F\'{\i}sica,
  Av. Fernando Ferrari, Goiabeiras, 
29075-910, Vit\'oria, ES, Brasil}

\affiliation{National Research Nuclear University “MEPhI”, Kashirskoe sh. 31, Moscow 115409, Russia }

\author{ Rudnei O. Ramos}
\affiliation{Universidade do Estado do Rio de Janeiro,  Departamento
  de F\'{\i}sica Te\'orica, 20550-013 Rio de Janeiro, RJ, Brazil}

\begin{abstract}

Cosmological scalar perturbation theory studied in the Newtonian gauge
depends on two potentials $\Phi$ and $\Psi$. In General Relativity
(GR) they must coincide ($\Phi=\Psi$) in the absence of anisotropic
stresses sourced by the energy momentum tensor. On the other hand, it
is widely accepted in the literature that potential deviations from GR
can be parameterized by $\Phi\neq \Psi$. The latter feature is
therefore present in both GR cosmologies equipped with shear viscous
fluids or modified gravity. We study the evolution of scalar matter
density perturbations using the redshift-space-distortion based
$f(z)\sigma_8(z)$ data as a tool to differentiate and characterize the
imprints of both scenarios. We show that in the $f(z)\sigma_8(z)$
evolution both scenarios yields to completely different imprints in
comparison to the standard cosmology. While the current available data
is not sensitive to distinguish modified gravity from viscous shear
cosmologies, future precise data can be used to break this
indistinguishability.

\end{abstract}

\maketitle

\section{Introduction}

The several available cosmological observables powerfully constrain
the background expansion of the universe as the one dictated by the
flat-$\Lambda$CDM model, i.e., a GR based description for gravity
composed of baryonic plus dark matter ($\Omega_{m0}\sim 0.3$) and a
cosmological constant $\Lambda$ $(\Omega_{\Lambda}\sim 0.7)$. However,
the background expansion, which can be characterized by the Hubble
term evolution $H_{\Lambda}$ in the $\Lambda$CDM model, can also be
achieved in modified gravity scenarios if suitable choices in their
degrees of freedom are made. Therefore, investigation of the
perturbative cosmological sector is necessary as an additional tool
such as to increase our ability to distinguish GR from its possible
candidate extensions.

The recent detection of gravitational waves from
$GW170817$~\cite{Monitor:2017mdv} has set the bound on the
gravitational wave speed $c_{gw}$ compared to the light speed $c$ as
$\left| c^2_{gw}/c^2 -1\right| < 5 \times 10^{-15}$. This result
severely reduces the available parameter space of generic
Lorentz-breaking modifications of gravity, as for example some
branches of the Horndeski (and Beyond-Horndeski)
theory~\cite{Creminelli:2017sry,Ezquiaga:2017ekz}. Hence, the
radiative sector of gravitational theories seems to be tightly close
to GR, but  the potential sector could still have space to manifest
some differences from the standard gravity. 

Using the Newtonian gauge for cosmological scalar perturbations in an
expanding, homogeneous and isotropic flat Universe, the line element
reads
\begin{equation}\label{metric}
ds^2 = a^2(\tau) \left[ -(1+2\Phi)d\tau^2 + (1-2\Psi)\delta_{ij}dx^i
  dx^j \right],
\end{equation}
where $\tau$ is the conformal time and $\Phi$ and $\Psi$ are the
metric perturbations. It is quite usual in the literature to
parameterize phenomenological departures from GR in terms of a
difference between the scalar potentials, $\Phi \neq \Psi$ (see, e.g.,
Ref.~\cite{Ade:2015rim} and references therein). 

Apart from the perspective given above, the issue we want to stress
out in this work is that $\Phi \neq \Psi$ is also naturally achieved
in GR cosmologies if the energy-momentum tensor $T^{\mu\nu}$ of some
of the energy components possesses anisotropic stresses like, e.g.,
shear viscosity.  Then, we are left with the question: Is the possible
inference of $\Phi \neq \Psi$ from observational data actually
indicating a manifestation of modified gravity or would it be due to
some non-conventional aspect of the universe's energy content? In
order to investigate this question we develop scalar perturbations in
two different cosmologies, namely, i) GR gravity equipped with a
cosmological constant and viscous (shear) matter and ii) modified
gravity theories via usual parameterizations of the Poisson equation.
Then, we compare the predictions for the growth of matter
perturbations via the redshift-space-distortion based
$f(z)\sigma_8(z)$ measurements~\cite{Song:2008qt}. In order to probe
only the perturbative sector of these two approaches we will assume
that both share the same background expansion as the one given by the
standard flat-$\Lambda$CDM model. To some extent, similar strategies
have been employed in Ref.~\cite{Ade:2015rim,Perenon:2015sla}.

Shear viscous effects in cosmology are in fact receiving interest in the recent literature as a possible way of understanding
different physical phenomena that might be in play both in the late
universe~\cite{Barbosa:2017ojt,Thomas:2016iav,Kopp:2016mhm,Anand:2017wsj}
as also in the early
universe~\cite{BasteroGil:2011xd,Bastero-Gil:2014jsa}. These recent
interests on shear viscous effects show that there are clear
motivations for a deeper study of their possible effects and relevance
in cosmology, which might eventually also provide relevant information
about the nature of the dark matter itself. In the present work, our
focus on the shear viscous effects is directly connected on how they
contribute at the perturbation level and the issue of having $\Phi
\neq \Psi$ for the gravitational perturbed potentials as a consequence
of the presence of a nonvanishing shear viscosity. Our main interest
then is to understand how this compares with the apparent similar
situation in the context of modified gravity models, quantifying the
possible differences in the two cases. 

Among all possible viscous effects that could also affect the cosmological expansion, bulk viscosity is also very representative in the literature \cite{Zimdahl:1996ka}. Here, we neglect this effect in a first approximation since it does not lead to $\Phi
\neq \Psi$ contribution and this could also add an undesirable degenerescence in the proposed comparison between viscous effects and modified gravity. Also, bulk viscosity modifies the background expansion. This would place some difficulties in the strategy we want to promote here since it would be impossible to set the same background evolution for both scenarios (viscous cosmology and modified gravity). However, in spite of the degeneracy in introducing bulk viscosity, kinetic pressure and baryons effects, we estimate their impact in our analysis in general lines. We reinforce such aspects with the discussion presented in section \ref{extra}.  

This paper is organized as follows. In section~\ref{sec2} we develop
the perturbation dynamics of the model with shear viscosity. {}For
this analysis, we take particular advantage of the results obtained in
Ref.~\cite{Barbosa:2017ojt}, where we have placed an upper bound on
the magnitude of dark matter shear viscosity allowed by the matter
clustering observations. In section~\ref{modgrav}, we develop the
perturbed scalar equations for the case of modified gravity and
present the parameterizations that will be used in this work. In
section~\ref{results}, we give our analysis of the quantitative
comparison between the GR plus shear viscosity case and contrast these
results with  the modified gravity one by making use of the evolution
of the $f\sigma_8$ observable. {}Finally, in section~\ref{conclusion}
we give our conclusions.

\section{Dynamics of the viscous (shear) dark matter model}
\label{sec2}

We start by focusing on the $\Lambda$CDM model and by assuming that
matter behaves as a viscous/dissipative component possessing shear
viscosity. This type of approach has been used already a number of
times in the recent literature (see, e.g.,
Refs.~\cite{Barbosa:2017ojt,Anand:2017wsj}).  The general structure of
this model is given by the field equations derived from GR,
\begin{equation}
R_{\mu \nu} - \frac{1}{2}g_{\mu \nu} R - \Lambda g_{\mu \nu} = 8\pi G
T_{\mu \nu},
\label{fieldeq}
\end{equation}
where $T_{\mu \nu}$ stands for the total energy momentum tensor of the
viscous matter. This tensor possesses the dissipative effect in the
form of shear viscosity such that~\cite{Landau:1971,Weinberg:1972}
\begin{equation}
T^{\mu \nu} = \rho_v u^{\mu} u^{\nu} - p_v \left( g^{\mu \nu} -
u^{\mu} u^{\nu} \right) + \Delta T^{\mu \nu},
\label{Tmunu}
\end{equation}
where the component $\Delta T^{\mu \nu}$ is the viscous contribution
to the fluid in the form of shear viscosity,
\begin{eqnarray}
\Delta T^{\mu \nu} &=&  \eta \left[ u^{\mu ; \nu} + u^{\nu  ; \mu}
  -u^{\rho} \nabla_{\rho} \left( u^{\mu} u^{\nu} \right)
  \right. \nonumber \\ &-& \left.  \frac{2}{3}   \left(  g^{\mu \nu} -
  u^{\mu} u^{\nu}\right) \nabla_\rho u^\rho \right],
\end{eqnarray}
and $\eta$ is the shear viscosity coefficient.
Although the above formulation represents a non-causal (Eckart) theory in hydrodynamics \cite{Landau:1971} it seems enough for the phenomenological applications we have in mind in this work. The coefficient of
shear viscosity, being a transport coefficient, is typically
proportional to the particle free mean path as in any microscopic
formulation of viscosity effects and it can also depend on the density
and temperature of the fluid. This, however, implies on the knowledge
of the microscopic physics of the interactions between the dark matter
particles. As we do not have in mind specific candidates for dark
matter particles, most of the time it is assumed for $\eta$ some
simple functional form in terms of the fluid density, like $\eta
\propto \rho_v^{\alpha}$~\cite{Barbosa:2017ojt}. This functional form
has the advantage of allowing for a completely model independent
analysis, where we do not need to specify properties related to the
dark matter fluid inherent to its microscopic physics.  Alternatively,
we can also see $\eta \propto \rho_v^{\alpha}$ as a consequence of
appropriately  choosing the dimensional scale as the fluid density
itself and where all microscopic dimensional parameters are expressed
in terms of this scale up to appropriate dimensionless constants.  In
the present work we will not be interested in these details and it
will suffice for our objectives of the comparison of the shear viscous
effects with those from modified gravity by simply adopting  $\eta$ to
be a constant parameter.  {}For simplicity, we will also set the dark
matter kinetic pressure to $p_v=0$. This guarantees a pressureless
matter fluid at the background level as in the standard
cosmology. Indeed, shear viscosity does not act at the background
level.

As already mentioned in the introduction, our starting point is based
on setting the line element of a Friedmann-Lemaitre-Robertson-Walker
(FLRW) expansion up to first order in scalar perturbation according to
Eq.~(\ref{metric}). Hence, from Eq.~(\ref{fieldeq}), the expansion
rate here follows the usual flat-$\Lambda$CDM model with
\begin{equation}
H^{2}= H^{2}_0 \left[\Omega_{v0}(1+z)^3+1-\Omega_{v0}\right]^{1/2},
\end{equation}
where the today's viscous matter density adopted is $\Omega_{v0}=8\pi
G \rho/ 3 H^2_0=0.3$. In the above equation the expansion rate is
written in the more familiar form as $H=\dot{a}/a$, where the symbol
dot ($^{.}$) represents the derivative with respect to the cosmic time
($t$). 

Our next step is to review the perturbed equations for shear viscous
cosmologies. We have developed it in great detail in
Ref.~\cite{Barbosa:2017ojt}, so below we only give the relevant
expressions needed for the present study. Applying Eq.~(\ref{metric})
to the Einstein equations we obtain, for example, the
$(0,0)$-component, in momentum space. It reads
\begin{equation}\label{Einstein0}
-k^2 \Psi - 3\mathcal{H} \left( \Psi^{\prime} +  \mathcal{H} \Phi
\right) = \frac{3}{2} \Omega_v \mathcal{H}_0^2 a^2 \Delta_v,
\end{equation}
while the $(0,i)$-component is given by
\begin{equation}\label{0i}
-k^2 \left( \Psi^{\prime} + \mathcal{H}\Phi \right) =
\frac{3}{2}\mathcal{H}_0^2 \Omega_v  a \theta_v,
\end{equation}
where $\mathcal{H}=\frac{a^{\prime}}{a}$, with the symbol "$\prime$ "
corresponding to a derivative with respect to the conformal time
($\tau$), $k$ is the (comoving) momentum and
$\Omega_v=\Omega_{v0}/a^3$.  When writing Eq.~(\ref{Einstein0}), we
have also used the definition of the density contrast,  $\Delta_v =
\delta \rho_v / \rho_v$.  {}From the ($0,i$)-component of the
Einstein's equation~(\ref{0i}), we obtain the definition for the
velocity potential $\theta= \partial_i \delta u^{i}$.  {}Finally, the
evolution of the perturbation potentials $\Psi$ and $\Phi$ are encoded
in the   $i - j$ component of the Einstein equation and given
explicitly by the expression
\begin{align}
& \left[ \Psi^{\prime \prime} + \mathcal{H} \left( 2\Psi + \Phi
    \right)^{\prime} +  \left( 2 \mathcal{H}^{\prime} + \mathcal{H}^2
    \right) \Phi + \frac{1}{2} \nabla \left( \Phi - \Psi \right)
    \right] \delta^i_j \nonumber \\  & - \frac{1}{2} \partial_i
  \partial_j  \left( \Phi - \Psi  \right) = 4 \pi G a^2 \delta T^i_j ,
\label{Eqij}
\end{align}
where the perturbed energy-momentum tensor is obtained from
Eq.~(\ref{Tmunu}), which gives
\begin{eqnarray}
\delta T^i_j &=& - \eta g^{ik} \delta^l_j \left( \delta u_{k,l} +
\delta u_{l,k} - \frac{2}{3}a^2 \delta u^m_{\, ,m} \delta_{kl}
\right),
\end{eqnarray}
and from the $i \neq j$ case of the above equation, we find that
Eq.~(\ref{Eqij}) corresponds to
\begin{equation}\label{Einstein1}
-\frac{k^2}{2}(\Phi - \Psi) = \frac{3\mathcal{H}^2}{\rho}\eta \,
\theta.
\end{equation}
Equation~(\ref{Einstein1}) makes it clear that when $\eta \neq 0$ we
have that $\Phi \neq \Psi$.  This demonstrates a notable feature of
the presence of the shear viscosity (anisotropic stress), i.e., the
Newtonian potentials do not coincide. It is worth noting that $\Phi
\neq \Psi$ is also seem in general in the literature as a
manifestation of modified gravity theories~\cite{Amendola:2007rr,
  MGpotentials1, MGpotentials2, MGpotentials3}.

By combining the above relations (the interested reader can also
consult Ref.~\cite{Barbosa:2017ojt} for further details), we obtain
\begin{eqnarray}\label{euler}
&&a^2 \frac{d^2 \Delta_v}{d a^2}+\left( 3 -\frac{3}{2} \Omega_v
  \frac{H_0^2}{H^2} +A + k^2 B \right) a \frac{d \Delta_v}{da}
  \nonumber \\ &-&\frac{3}{2} {\Omega_{v}} \frac{H_0^2}{H^2} \Delta_v
  = 0,
\end{eqnarray}
where the factors $A$ and $B$ appearing in the above equation  are
defined, respectively, as
\begin{eqnarray}
A &=&  \frac{2\tilde{\eta}}{3 \Omega_v^2} \frac{H}{H_0},
\label{Aeta}
\\ \nonumber \\ B &=&   \frac{4 \tilde{\eta}}{27 a^2 \Omega_v H H_0},
\label{Beta}
\end{eqnarray}
where $\tilde{\eta}=24 \pi G \eta/H_0$ is the dimensionless shear viscous parameter. One can see explicitly that shear viscosity leads to contributions to
the Hubble friction term in the differential equation for the matter
density contrast. 

It is worth noting that the quantity $\Delta_v$ introduced above corresponds to the density contrast of the total viscous matter. The correspondent quantity in the $\Lambda$CDM model (let us say $\Delta_m$) is obtained with $\tilde{\eta}=0$. In the latter model, at linear order, the baryonic perturbations $\Delta_b$ follow the evolution of the CDM ones $\Delta_{CDM}$. Consequently, large scale structure observables like the growth rate used below in this work is sensitive to $\Delta_m$ rather than either the density contrast in the cold dark matter or the density contrast of the baryonic or a combination of both. Even if we promote a split of the total viscous matter perturbation $\Delta_v$ into the "viscous`` dark component (let us say $\Delta_{vCDM}$) and the baryonic one, there is only a slighlty difference between $\Delta_{v}$ and $\Delta_{vCDM}$ as shown in \cite{Barbosa:2017ojt}. Nevertheless, the above arguments hold only for the linear regime. At the nonlinear regime baryons present much more friction (and, consequently, viscosity) than CDM as revealled by the Bullet cluster \cite{Clowe:2003tk, Clowe:2006eq}.

\section{Modified gravity at linear perturbative level}
\label{modgrav}

In the previous section we have obtained the equations for the case in
which shear viscosity sets the magnitude of the inequality $\Phi \neq
\Psi$ via Eq.~(\ref{Einstein1}). Let us now see in this section how
the effects of modified gravity can also be parameterized by
differences between $\Phi$ and $\Psi$.  In particular, we want to
explore the consequences of choosing the usual parameterizations of
modified gravity in which the {\it slip} parameter, defined by the
ratio $\Psi/\Phi$, is used to quantify deviations from GR. Then, we
assume that the phenomenology for dealing with modifications of
gravity at cosmological scales merely sets the inequality
\begin{equation}
\Phi \neq \Psi \,\,({\rm Modified\,\,Gravity}).
\end{equation}

Even when adopting modifications of gravity, we assume that the theory
is conservative and that the usual conservation laws apply, i.e.,
$\nabla_{\mu} T^{\mu\nu}=0$. Our approach for dealing with scalar
perturbations in a parameterized modified gravity theory consists in
combining the perturbed continuity equation for the density contrast
$\Delta$ for a pressureless fluid and in the modified gravity context,
given by
\begin{equation}
\Delta^{\prime}+a \theta = 0,
\label{Deltaeq}
\end{equation}
with the Euler equation
\begin{equation}
(a\theta)^{\prime} + \mathcal{H} a\theta - k^2 \Phi = 0,
\label{thetaeq}
\end{equation} 
to obtain the result
\begin{equation}
\Delta^{\prime\prime} + \mathcal{H} \Delta^{\prime}+ k^{2} \Phi=0.
\label{genDelta}
\end{equation}
Therefore, Eq.~(\ref{genDelta}) tell us that the matter clustering
growth (the observable we are interested in) depends only on the
potential $\Phi$. However, on sub-horizon scales we can also write
down the Poisson equation as
\begin{equation}
-k^{2} \Psi = \frac{3}{2}\Omega_{m}\mathcal{H}^{2}_0 a^{2} \Delta.
\label{genPoisson}
\end{equation}
At this point, it is worth noting that the standard equation for the
evolution of $\Delta$ is obtained by assuming $\Phi=\Psi$ and
combining the above two equations. It is exactly this step we want to
avoid. Instead, we will adopt typical parameterizations of
Eq.~(\ref{genPoisson}) found in the literature to explore the
phenomenology of modified gravity. The some possible choices for the
functional parameterization that we will use in our study is explained
next.

Equations (\ref{genDelta}) and (\ref{genPoisson}) involve three
different functions. Indeed, if anisotropic stresses are neglected in
the energy-momentum tensor one obtains $\Phi=\Psi$ and a homogeneous
second order differential equation for $\Delta$ is
obtained. Departures from the standard model, i.e., the $\Lambda$CDM
model, are usually parameterized in the literature by $\Phi\neq\Psi$.
Since we want to investigate small deviations from GR, which are
relevant for the structure formation process, we will then follow an
analogous strategy as used, e.g., in Refs.~\cite{Perenon:2015sla,
  Ade:2015rim} and set the background evolution to be the same as in
$\Lambda$CDM.  As in Ref.~\cite{Ade:2015rim}, we adopt a Poisson type
equation for the potential $\Phi$, such that
\begin{equation}\label{poissonmodified}
-k^2 \Phi \equiv 4\pi Ga^2 \mu(a,k)\rho \Delta,
\end{equation}
where $\mu(a,k)$, sometimes also denoted by the function $Y(a,k)$,
incorporates to the relativistic Poisson equation possible
contributions from clustering dark energy.

Combining Eq. (\ref{genDelta}) with the parameterization
Eq. (\ref{poissonmodified}) and using the scale factor as the
dynamical variable we obtain the following equation for the evolution
of the matter density contrast
\begin{eqnarray}
&& a^2 \frac{d^2 \Delta}{da^2} + \left( 3 + \frac{a}{H}\frac{dH}{da}
  \right) \frac{d\Delta}{da} \nonumber \\ && - \frac{3}{2}\Omega_m
  \frac{H_0^2}{H^2} \mu(a,k) \Delta = 0.
\label{constrastedensidade}
\end{eqnarray}
The function $\mu(a,k)$ can in principle depend on time (here given in
terms of the scale factor $a$ dependence) and the scale (via the
wavenumber-mode $k$). 

By comparing Eq. (\ref{euler}) with Eq. (\ref{constrastedensidade})
one realizes one important difference between the shear viscous
scenario and modified gravity. Shear viscosity acts by damping the
Hubble friction term in Eq. (\ref{euler}). This conclusion is in
agreement with the recent study performed in
Ref.~\cite{Kremer:2018kul}, which also shows how shear viscosity damps
the growth of structures. It is worth noting that such damping is not
present in modified gravity scenarios. In fact, for example in
$f(R)$-type of models for modified gravity, the resulting effect is
rather usually associated with the boosting of the agglomeration
rate~\cite{Alam:2015rsa} and of the matter power
spectrum~\cite{Pogosian:2010tj}.

According to Ref.~\cite{Ade:2015rim}, one possible way to employ the
parameterization  in the form as given in Eq.~(\ref{poissonmodified}),
occurs by choosing the function $\mu(a,k)$ as 
\begin{equation}
\mu(a,k) = 1 + f(a)\frac{1 + c(\lambda H/k)^2}{1+(\lambda H/k)^2 },
\label{21}
\end{equation}
where $c$ and $\lambda$ are constant parameters. {}For the sake of
simplicity and without loss of generality we will fix $\lambda=1$. At
small scales (large $k$), $\mu \rightarrow 1 + f(a)$, while for large
scales (small $k$), $\mu \rightarrow 1 + f(a) c$. Then, in practice
the scale dependence plays no decisive role for astrophysical
applications we have in mind.  Thus, we proceed now by adopting the
following simpler structure,
\begin{equation}
\mu_1(a) = 1+ E_1 \frac{H^2_0}{H^2},
\label{mu1}\end{equation}
where $E_1$ is a constant parameter, with the parameter $c$ absorbed
in the definition of $E_1$.  Equation~(\ref{mu1}) will be the first
parameterization form we will use. {}For completeness, we will also
use two more and that will be defined below.

The range of values of the parameters presented in the Eq.~(\ref{21})
depends on the modified gravity theory. {}For instance, in the case of
the $f(R)$ theories, with a chameleon mechanism, the coefficients are
positive, implying that the gravitational coupling is enhanced
compared with the GR case~\cite{Alam:2015rsa,Pogosian:2010tj}. An
enhanced gravitational coupling leads to a stronger matter
agglomeration. Even if such property of modified gravity theories must
be verified case by case, it remains a quite general feature, at least
to our knowledge. {}For this reason, we will also consider $E_1$ as a
positive quantity. Hence, it is already possible, at this level, to
predict that modified gravity acts on matter agglomeration in the
opposite sense compared to shear viscosity: While the shear viscosity
suppresses the matter agglomeration, modified gravity acts mainly in
the sense of enhancing the formation of structures.

Besides the parameterization given by Eq.~(\ref{mu1}), we will also
make use of two more that are conventionally considered in the
literature. More specifically, we also consider the parameterization
according to the proposal of Ref.~\cite{Bean:2010zq} and define
\begin{equation}
\mu_2(a) = 1 +( E_2 e^{-\frac{k}{k_c}}-1),
\label{mu2}\end{equation}
where the scale $k=0.1 h Mpc^{-1}$ has been fixed. Indeed, it is a
sub-horizon mode and still linear at $a_0$. The free constant
parameters are $E_2$ and $k_c$. The GR limit occurs for $E_2 =1$ and
$k_c \rightarrow\infty$. 

{}Finally, we also consider the parameterization proposed in
Ref.~\cite{Amendola:2007rr} and studied recently by the authors of
Ref.~\cite{Resco:2017jky}, given by
\begin{equation}
\mu_3(a) = 1 + E_3 \frac{2\left[1 +
    2\Omega_m(a)^2\right]}{3\left[1+\Omega_m(a)^2\right]},
\label{mu3}\end{equation}
and which is inspired within the DGP gravity scenario~\cite{DGP}.

\section{Results}
\label{results}
\begin{center}
\begin{figure}[!htb]
\subfigure[]{\includegraphics[width=8cm]{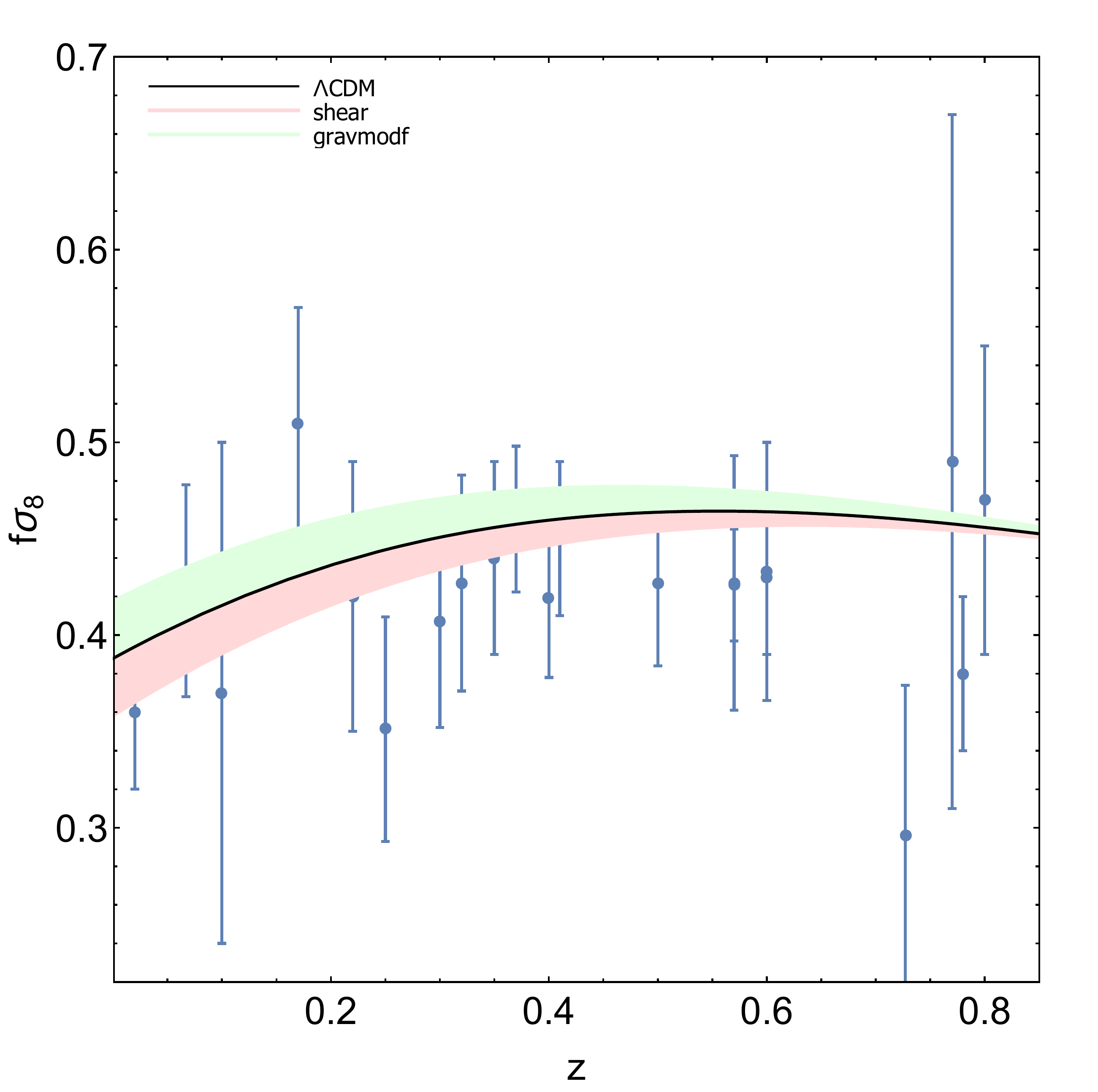}}
\subfigure[]{\includegraphics[width=8cm]{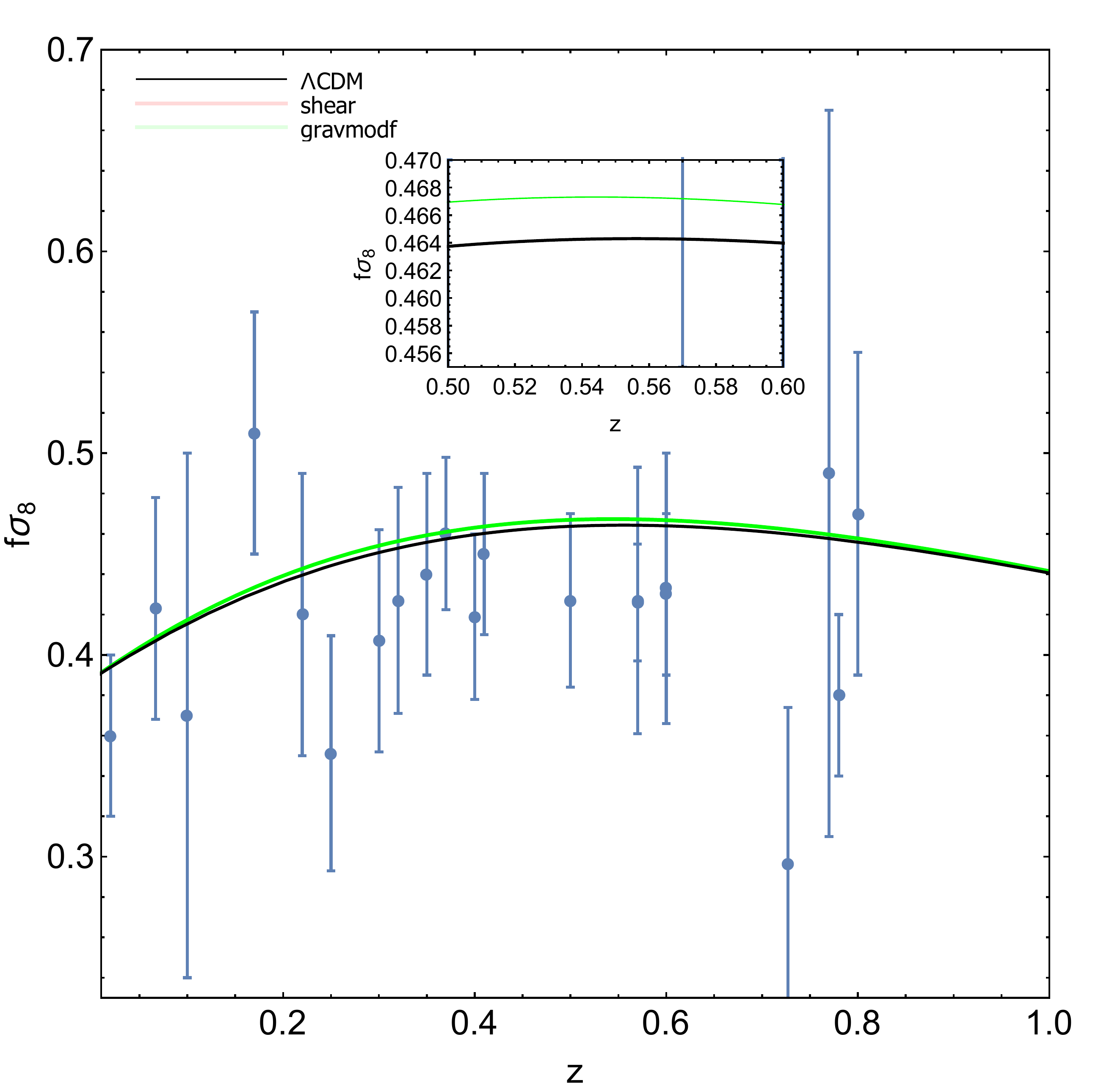}}
\caption{The $f \sigma_8$ observable as a function of the
  redshift. (a) {}For the shear viscous model the light-red area
  corresponds to the range of the viscous parameter $ 0 \leq
  \tilde{\eta}_0 \leq 2.593 \times 10^{-6}$. The green region shows
  the behavior of modified gravity model with the $\mu_1$
  parameterization with the range $0 \leq E_1 \leq 0.225$. (b) Plot of
  the difference between top green line and bottom red line from
  (a). At $z=0.54$ we have the largest difference between shear and
  the modified gravity model (detail shown in the inset).}
\label{Fig1}
\end{figure}
\end{center}

\begin{center}
\begin{figure}[!htb]
\subfigure[]{\includegraphics[width=8cm]{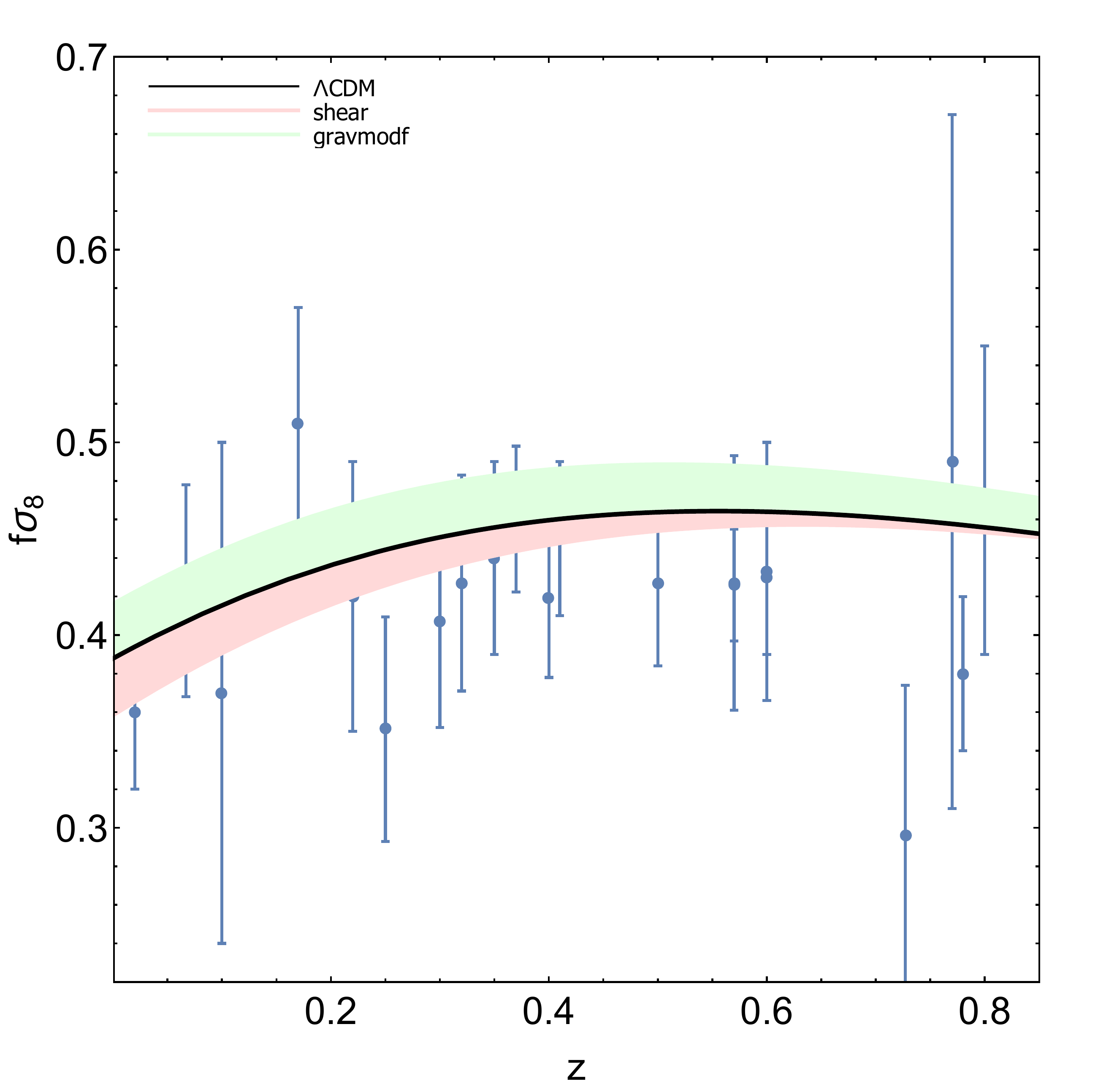}}
\subfigure[]{\includegraphics[width=8cm]{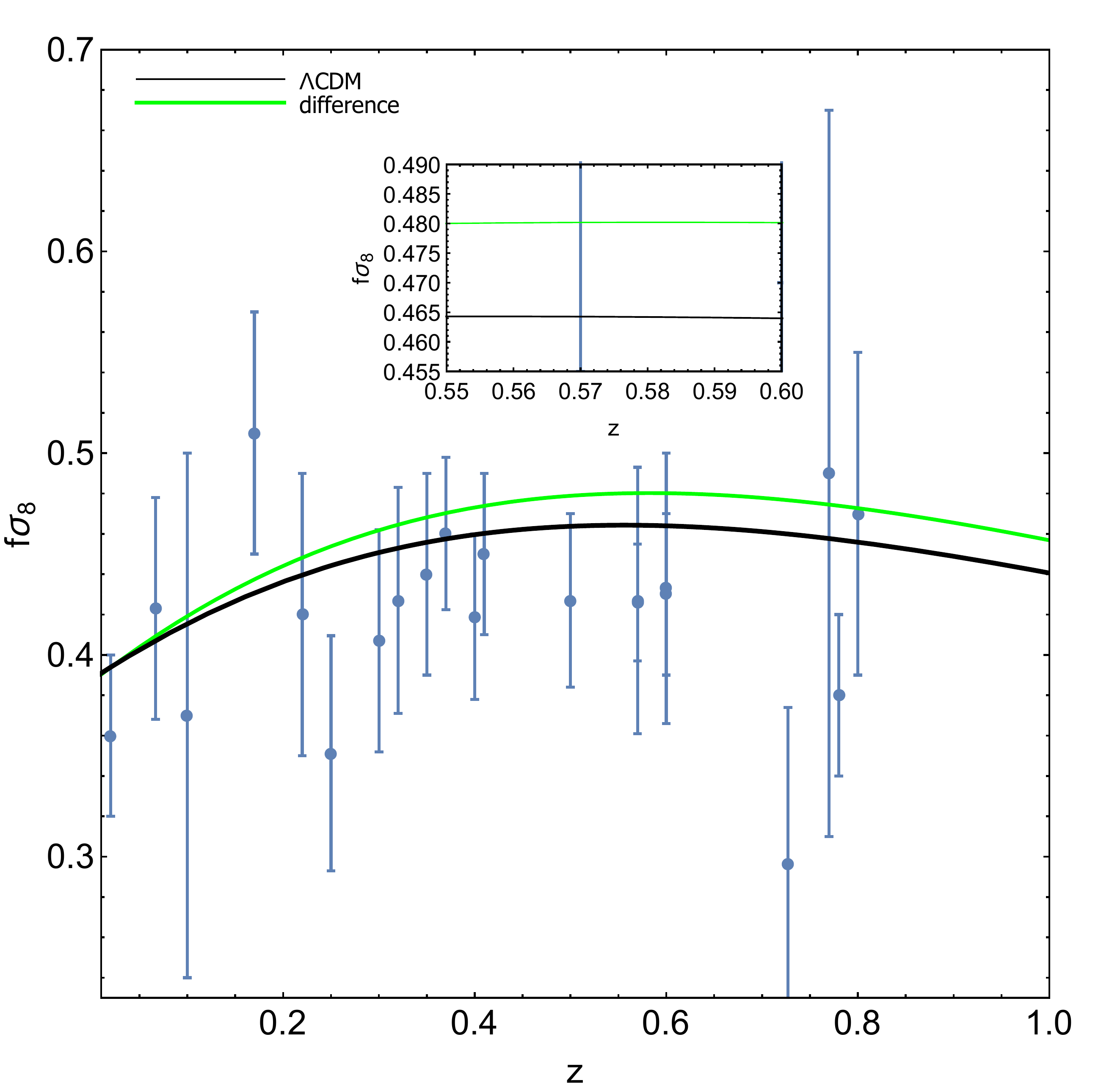}}
\caption{The $f \sigma_8$ observable as a function of the
  redshift. (a) {}For the shear viscous model the red lines correspond
  to the range of the viscous parameter $ 0 \leq \tilde{\eta}_0 \leq
  2.593 \times 10^{-6}$. Green lines shown the behavior of modified
  gravity models with the $\mu_2$ parameterization. Different values
  for $E_2$ parameter $1.1 \leq E_2 \leq 1.23$ and $1 \leq k_c \leq
  1.1$. (b) Plot of the difference between top green line and bottom
  red line, from (a). At $z=0.58$ we have the biggest difference
  between shear and the modified gravity model (detail shown in the
  inset).}
\label{Fig2}
\end{figure}
\end{center}

{}From the three forms of parameterizations, given by
Eqs.~(\ref{mu1}), (\ref{mu2}) and (\ref{mu3}), respectively, we apply
them to  Eq.~(\ref{constrastedensidade}). The resulting equation for
each case can then be solved numerically for the density contrast
$\Delta$. Having also the result for the density contrast from the
shear viscous case and obtained from Eq.~(\ref{euler}), we can
calculate the growth function $f(z)$  for all these different
cases. The growth function $f(z)$ is defined as 
\begin{equation}
f(z)\equiv \frac{d\ln\frac{\Delta(a)}{\Delta(a_0)}}{d\ln
  a}=-(1+z)\frac{d\ln \frac{\Delta(z)}{\Delta(z_0)}}{dz},
\end{equation}
with $z=1/a-1$ and 
\begin{equation}
\sigma_8(z)=\sigma_{8}(z_0)\frac{\Delta(z)}{\Delta(z_0)},
\end{equation} 
is the redshift-dependent root-mean-square mass fluctuation in spheres
with radius $8h^{-1}$ Mpc.  The today's scale factor is set to unity,
$a_0=1$, thus, $z_0=0$. The today's value adopted here for the
variance of the density field at $z_0$ is $\sigma_{8}(z_0)=0.8$, which
is consistent with current observations.

Let us consider the results obtained by using the first
parametrization given by Eq.~(\ref{mu1}).  In {}Fig.~\ref{Fig1}a we
show the $f \sigma_8$ observable as a function of the redshift. The
light-red filled area corresponds to the shear viscous model. This
region is set by using our previous results from
Ref.~\cite{Barbosa:2017ojt} and corresponds to the range of the
viscosity parameter $ 0 \leq \tilde{\eta} \leq 2.593 \times 10^{-6}$
at $2 \sigma$ of statistical confidence level obtained in that
reference. Here, for convenience, we recall we have defined the dimensionless
viscous parameter $\tilde{\eta}=24 \pi G \eta/H_0$ and $\eta$ is
assumed to be a constant value, as we have already explained in
Sec.~\ref{sec2}.  The viscous shear model equals the $\Lambda$CDM
(black line) curve for the case of vanishing viscosity,
$\tilde{\eta}=0$. The viscosity parameter, being physically a
transport coefficient, should assume only positive values. Thus, its
effect acts smoothing the matter clustering in comparison to the
standard cosmology, which corresponds to the region below the black
line. The value $\tilde{\eta} = 2.593 \times 10^{-6}$ is the maximum
viscosity allowed by the available 21 data points shown in this figure
at $2\sigma$ of statistical confidence level (see
Ref.~\cite{Barbosa:2017ojt} for details). The shear model with
$\tilde{\eta} = 2.593 \times 10^{-6}$ is the lowest light-red line
plotted in {}Fig.~\ref{Fig1}a.  The green filled area corresponds to
modified gravity models based on $\mu_1$ and defined in
Eq. ~(\ref{mu1}). Since we expect the values for $\mu_1$ to be such
that they increase the intensity of gravity~\cite{Ade:2015rim}, then
$\mu_1$ only assumes positive values.  The consequence of this
imposition can be seen in {}Fig.~\ref{Fig1}a.  The green lines always
stay {\it above} the $\Lambda$CDM line, while the red lines,
corresponding to the shear viscosity effect, always stay {\it below}
the $\Lambda$CDM line.

It is worth noting that both models share the same asymptotic behavior
for high redshifts.  In particular, the value $E_1=0$ corresponds to
the $\Lambda$CDM model.  Having the bound on $\tilde{\eta}_0$ given
above in mind, we have plotted the green region  in {}Fig.~\ref{Fig1}a
according to the following criteria: We limit the maximum $f\sigma_8
(a=1)$ given by the modified gravity model to yield the same departure
in magnitude from the $\Lambda$CDM model, but in the opposite
direction, in comparison to the shear model. Then, if combining both
effects they approximately compensate the effect of each other both
today and in the asymptotic past at high redshifts. The combination of
both effects is seen in  {}Fig.~\ref{Fig1}b. Although very tiny, the
region around $z=0.54$ is where one finds the largest difference
between both effects (inset plot). 

In {}Fig.~\ref{Fig2} we present the results obtained by using the
second parameterization for the modified gravity effects and given by
Eq.~(\ref{mu2}). The color scheme follows the same as the one used in
{}Fig.~\ref{Fig1}. We notice from {}Fig.~\ref{Fig2}a that now the
modified gravity results spread at an uniform distance above the
$\Lambda$CDM result for a given value of the constant $E_2$. In
particular, at low redshifts we again observe a compensation of the
modified gravity effect by the shear viscous (or vice-versa), as is
apparent from  {}Fig.~\ref{Fig2}b, where we plot the difference
between the maximum differences for each case with respect to the
$\Lambda$CDM result. However, at high redshifts the difference starts
to get more and more appreciable. 

\begin{center}
\begin{figure}[!htb]
\subfigure[]{\includegraphics[width=8cm]{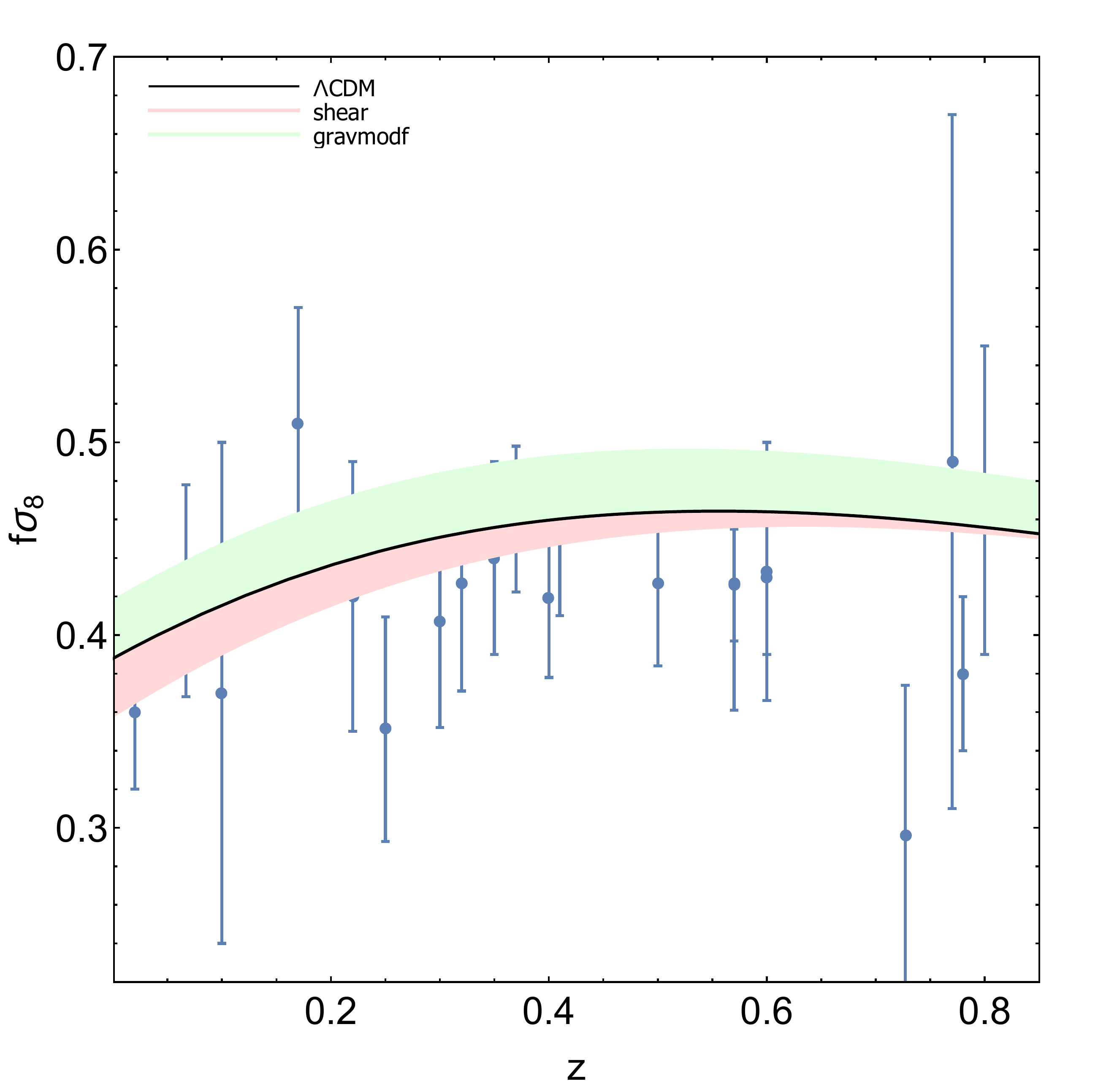}}
\subfigure[]{\includegraphics[width=8cm]{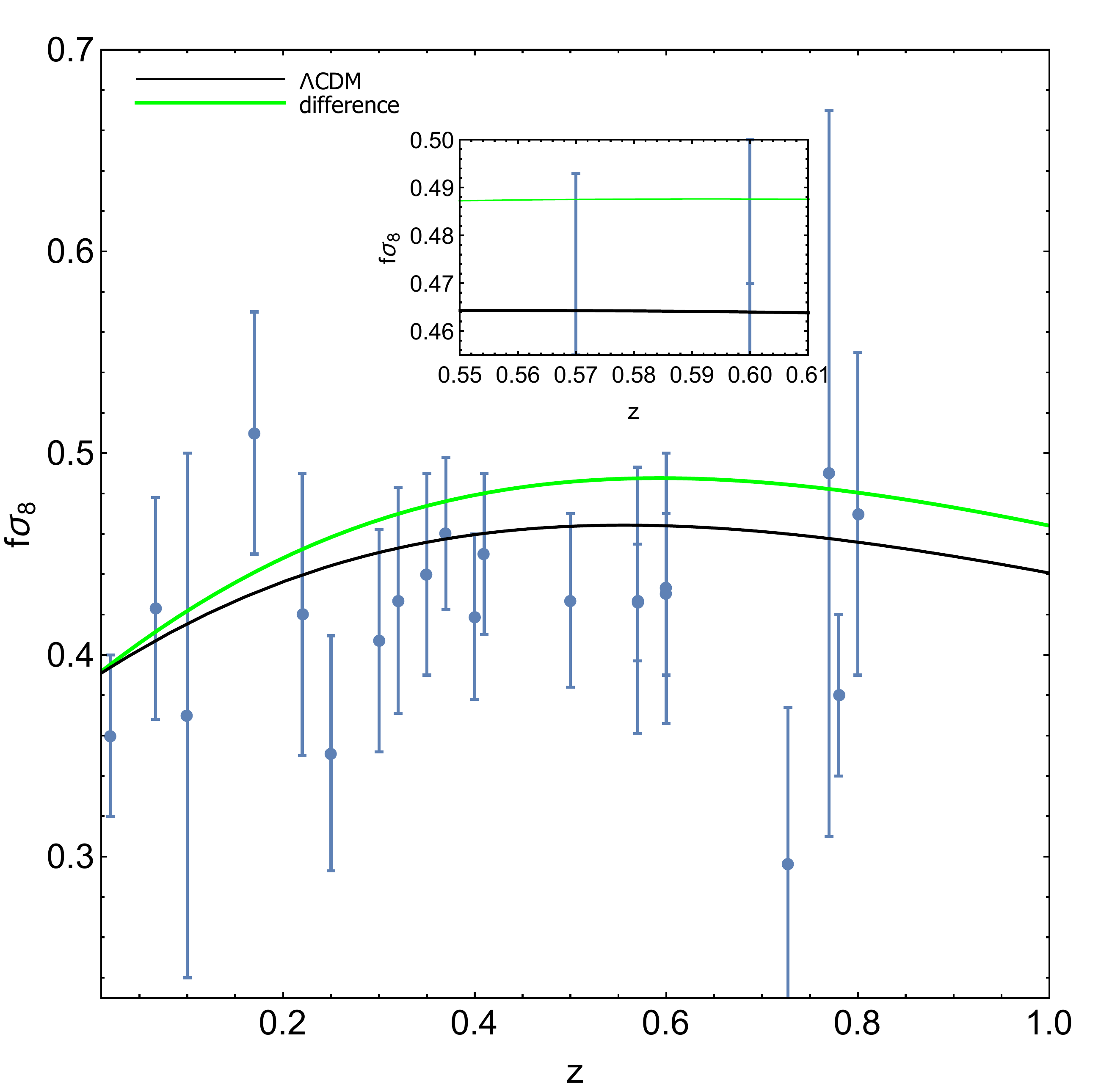}}
\caption{The $f \sigma_8$ observable as a function of the
  redshift. (a) {}For the shear viscous model the blue lines
  correspond to the range of the viscous parameter $ 0 \leq
  \tilde{\eta}_0 \leq 2.593 \times 10^{-6}$. Green lines shown the
  behavior of modified gravity models with the $\mu_3$
  parameterization. Different values for $E_3$ parameter $0 \leq E_3
  \leq 0.13$. (b) Plot of the difference between top green line and
  bottom red line, from (a). At $z=0.59$ we have the biggest
  difference between shear and the modified gravity model (detail
  shown in the inset).}
\label{Fig3}
\end{figure}
\end{center}

{}Finally, in {}Fig.~\ref{Fig3} we present the results obtained when
considering the parameterization given by Eq.~(\ref{mu3}). Once again,
the color scheme used in {}Fig.~\ref{Fig3}a follows the same as the
one already used in the previous two figures. The trend observed is
similar to the one obtained from the  parameterization given by
Eq.~(\ref{mu2}), where we have a tendency of shear viscous effects
maskering the modified gravity one and vice-versa at low redshifts,
but the difference increases more appreciably at high redshifts, as
seen in {}Fig.~\ref{Fig3}b.

\section{On the effects of baryons, bulk viscosity and other possible contributions}
\label{extra}

We have analysed so far the direct relation between shear viscosity and the slip parameter via the effects on the growth of matter scalar perturbations. This means we have ignored other possible hydrodynamical effects like the presence of a kinetic pressure and bulk viscosity and also the inclusion of a separated baryonic component. Now, our aim in this section is to include such effects in our discussion in order to set a rough estimation on the validity of our approach and showing how they also impact the matter clustering. This analysis shows therefore other degeneracy sources.

\begin{center}
\begin{figure}[!htb]
\subfigure[Results when the bulk viscosity is absent.]{\includegraphics[width=7.5cm]{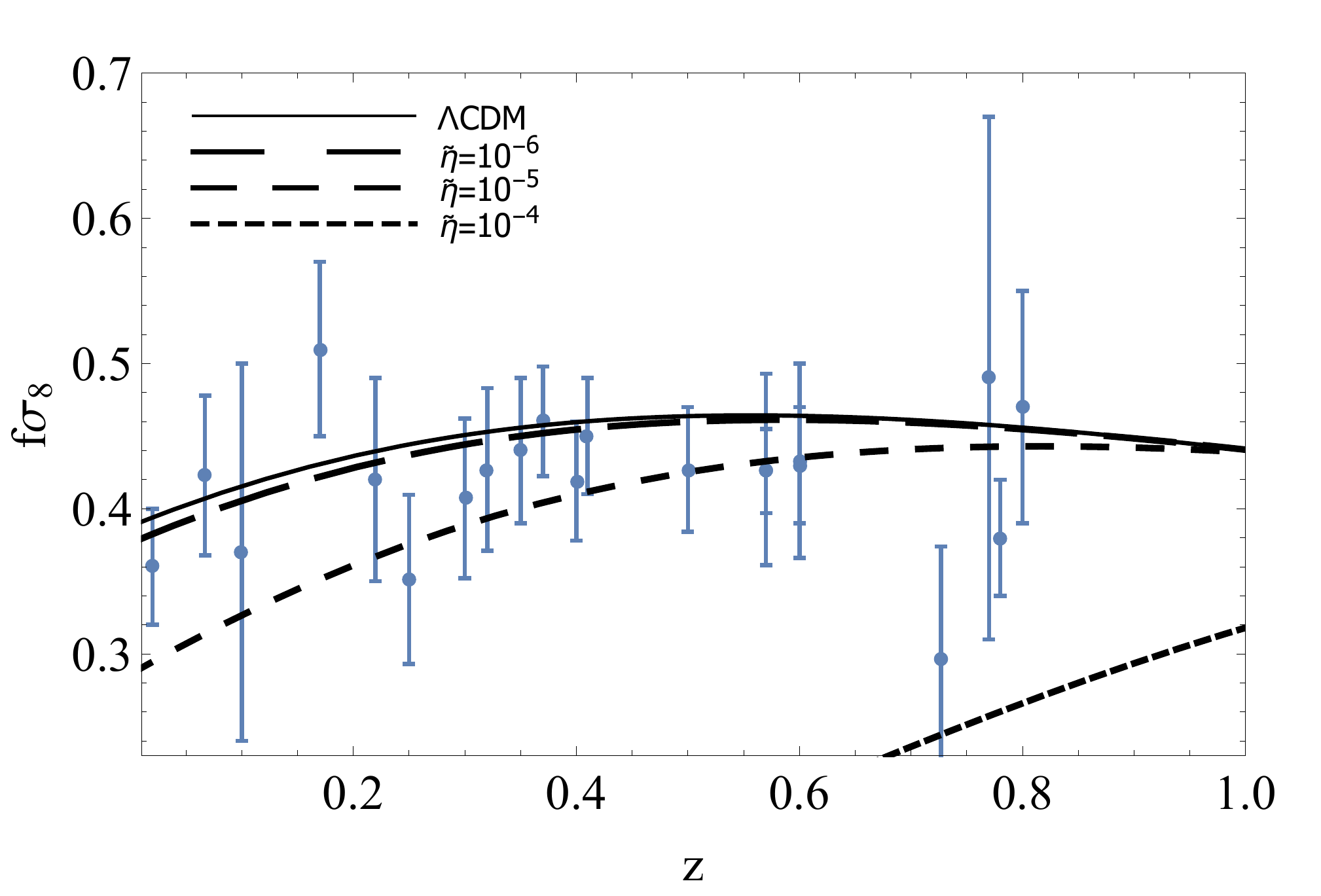}}
\subfigure[Results when the shear viscosity is absent.]{\includegraphics[width=7.5cm]{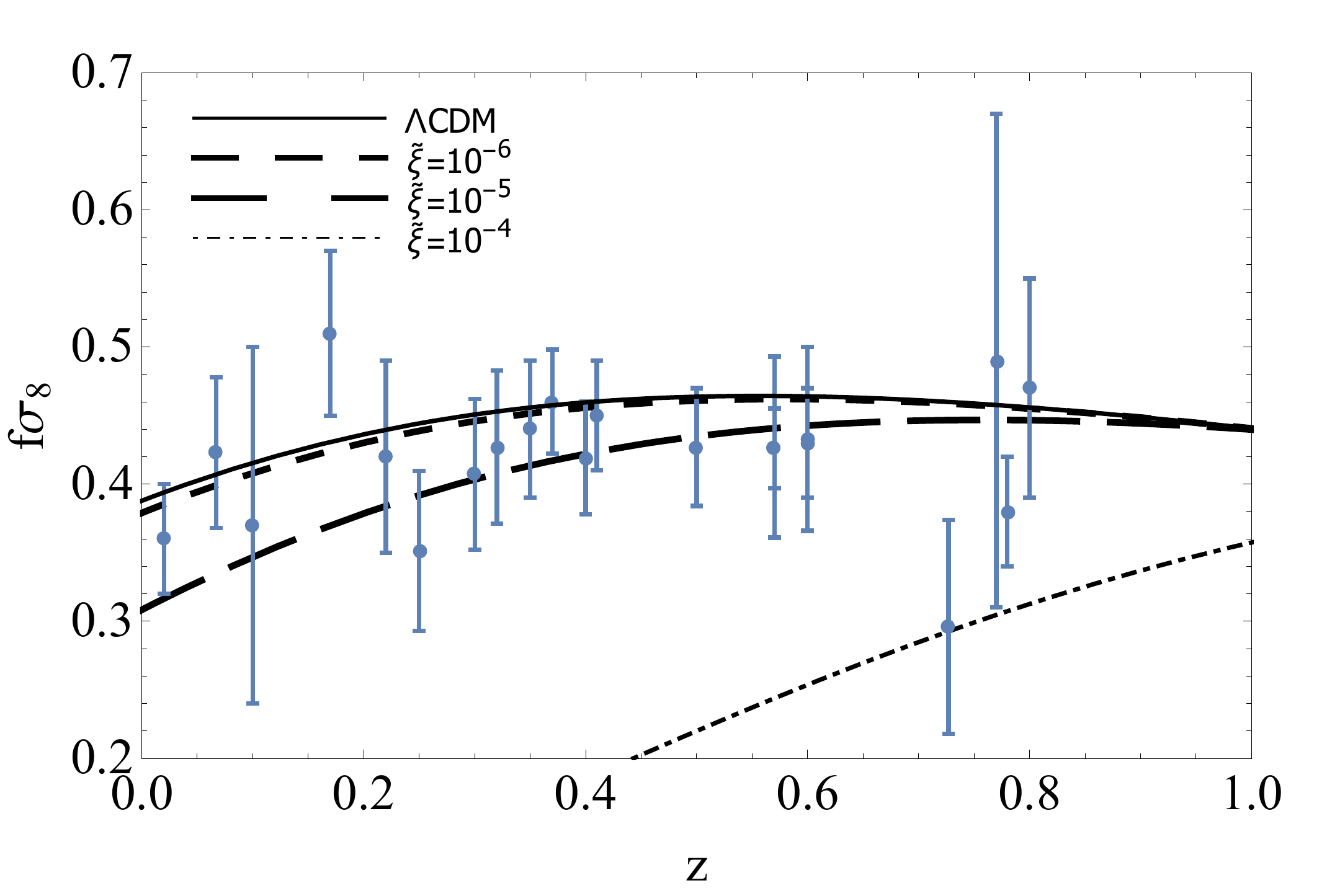}}
\caption{The linear growth against
the $f \sigma_8$ data as a function of the redshift in the absence and presence of the viscosities.}
\label{k01nulo}
\end{figure}
\end{center}

Since we have used shear viscosity in this work it is important to mention other dissipative properties. For example, bulk viscosity yields to an additional pressure at the background level. In a FLRW background, with expansion scalar $3H$, the bulk viscous pressure becomes $\Pi=-3H \xi$ where $\xi$ is the coefficient of bulk viscosity. Then, the total pressure of the fluid becomes $P=P_k+\Pi$ where $P_k$ is the kinetic pressure. The effective equation of state parameter can be written as
\begin{equation}
w=P/\rho=w_k-\tilde{\xi}/3,
\label{wtotal}\end{equation} 
where we have defined the dimensionless bulk viscous parameter $\tilde{\xi}=24 \pi G \xi / H_0$ and the kinetic pressure equation of state parameter $w_k=P_k/\rho$. As shown in Ref.~\cite{Barbosa:2017ojt} bulk and shear viscosities impact the growth of structures at the same level. Indeed, this happens only due to the perturbative dynamics features since values of order $\mathcal{\tilde{\xi}} \lesssim 10^{-5}$ would not affect the background scaling of the matter component. In practise, for such values of the bulk viscous parameter there is no impact at the background level. Therefore, in case the analysis performed in this work had taken also into account bulk viscosity the matter clustering would be even more suppressed. Since the slip parameter is then only related to the shear viscosity this means that in case both shear and bulk viscosities operate the bound on the slip parameter established before would be affected by a factor $\sim 2$. In order to demonstrate this results we present {}Fig.~\ref{k01nulo} from Ref.~\cite{Barbosa:2017ojt}. In the  first panel of {}Fig.~\ref{k01nulo}, we have only the effect of the shear viscosity, but no bulk viscosity. This is the same situation as previously shown in this work. In the second panel we show only the effect of the bulk viscosity. 

Now, concerning the possible impact of an extra baryonic component we also take advantage here of the discussion previously presented in Ref.~\cite{Barbosa:2017ojt}. It turns out that an extra baryonic fluid would contribute with a energy density $\rho_b$. Its corresponding perturbation $\Delta_b$ would act as a source term to the right hand side of
Eq. (\ref{Einstein0}). Also, there exist in this case separated conservation equations for the baryonic perturbations similarly to Eqs.~(\ref{Deltaeq}) and (\ref{thetaeq}). 

Let us now present equations in which we can compute both the evolution (obtained already in \cite{Barbosa:2017ojt}) of the perturbations of the viscous dark matter fluid (possessing both bulk and shear viscosities) as well as the perturbations of baryons. They are written according to
\begin{align}
& a^2 \frac{d^2 \Delta_b}{da^2} + \Big{\lbrace} 3 - \frac{3}{2} \frac{H_0^2}{H^2} \left[ \Omega_v (1+\omega_v) + \frac{\Omega_{b0}}{a^3} \right] \Big{\rbrace} a  -\frac{d\Delta_b}{da} \nonumber \\
&- \frac{3}{2} \frac{H_0^2}{H^2}\frac{\Omega_{b0}}{a^3}\Delta_b  = \Big{[} \frac{3}{2}\frac{H_0^2}{H^2} \Omega_v \nonumber \\
& + \frac{2 \tilde{\eta} a}{3 H_0 \Omega_v (1+2\omega _v)} 
\Big{(} \frac{3 H \omega_v}{a} + \frac{H^2}{H_0^2} \frac{\tilde{\xi } \nu}{\Omega_v} \left(\frac{\Omega_v}{\Omega_{v0}}\right)^{\nu} \Big{)} \Big{]} \Delta_v 
\nonumber \\ 
& - \frac{2 \tilde{\eta} H a}{3 H_0 \Omega_v (1+2\omega)} \frac{d\Delta_v}{da},
\label{Deltab}
\end{align}
where $\xi=\xi_0 (\Omega_v/\Omega_{v0})^{\nu}$ and $\eta =\eta_0 (\Omega_v/\Omega_{v0})^{\lambda}$.

The viscous fluid density perturbation equation 
is also modified when including baryons and it now becomes  
\begin{align}\label{eulerv}
&a^2 \frac{d^2 \Delta_v}{d a^2}+\left[3 - \frac{3}{2}\Omega_v \frac{H_0^2}{H^2} - \frac{3}{2}\frac{\Omega_{b0}}{a^3} \frac{H_0^2}{H^2} +\bar{A} +
    k^2 B \right] a \frac{d \Delta_v}{da} \nonumber \\ & + \left( \bar{C} + k^2 D \right) \Delta_v = \frac{3}{2} \frac{H_0^2}{H^2} \frac{\Omega_{b0}}{a^3} \frac{(1+2\omega_v)}{(1+\omega_v)}\Delta_b,
\end{align}
with $\Delta_v\equiv \delta\rho_v/(\rho_v+\rho_b)$ and where the factors $\bar{A}$, $B$, $\bar{C}$ and $D$ are defined, respectively, as

\begin{eqnarray}
\bar{A} &=& A +   \frac{3 \omega_v}{2(1+2\omega_v)(1+\omega_v)} 
\frac{\Omega_{b0}}{a^3}\frac{H_0^2}{H^2},
\label{newtermA}
\end{eqnarray}

\begin{eqnarray}
B = -\frac{w_v (1+\frac{4}{3}R)}{3 H^2 a^2 (1+ w_v)},
\label{newtermB}
\end{eqnarray}

\begin{eqnarray}
\bar{C} &=& C + \frac{9 \omega_v (2+6\omega_v + 5\omega_v^2)}{2(1+2\omega_v)(1+\omega_v)} 
\frac{\Omega_{b0}}{a^3}\frac{H_0^2}{H^2}. 
\end{eqnarray}
and

\begin{eqnarray}
D = \frac{ w_v^2 (1+\frac{4}{3}R)}{ H^2 a^2 (1+ w_v)}(1-\nu) +
  \frac{\nu \omega_v \left( 1 + 2w_v \right)}{1+w_v}\left(
  \frac{\Omega_v}{\Omega_{v0}} \right)^\nu.
\label{termD}  
\end{eqnarray}
The functions $A$ and $B$ have been defined in  \cite{Barbosa:2017ojt}.

In the above equations we have also introduced the quantity $R\equiv\tilde{\eta}/\tilde{\xi}$,
i.e., the ratio between the (dimensionless) shear and bulk viscosities which can also be explicitly written as
\begin{equation}
R = \frac{{\tilde \eta}_0}{{\tilde \xi}_0} \left(\frac{\Omega_v}{\Omega_{v0}}\right)^{\lambda-\nu}.
\end{equation}

We have now a two-fluid system described by the coupled equations 
(\ref{Deltab}) and (\ref{eulerv}) and where the baryon density contrast 
enters as a source term in the dark matter viscous equation one.

It is worth noting that a cosmological observable like $f\sigma_8$ takes into account the total matter. This is the case of the standard model in which concerning the linear perturbations both dark matter and baryons are treated as a single matter fluid. There is no distinction between them. Back to the possibility of a separated baryonic fluid let us define then an effective density contrast
\begin{equation}
\Delta_{eff}=\frac{\Omega_v \Delta_v+\Omega_b \Delta_b  }{\Omega_v +\Omega_b}
\label{deltaeff}
\end{equation}    
which would be used rather than $\Delta$.    

\begin{center}
\begin{figure}[!htb]
\includegraphics[width=8cm]{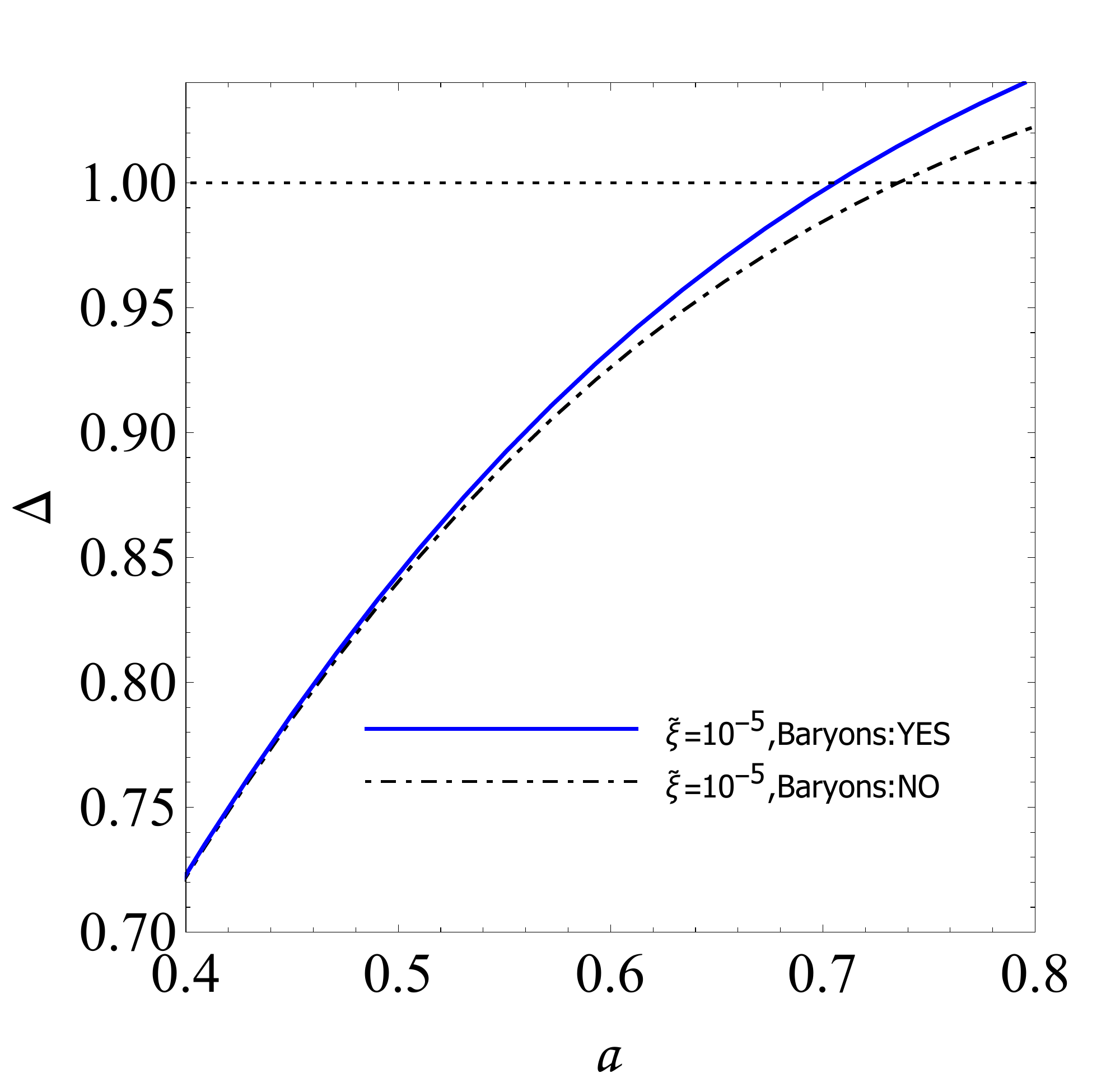}
\caption{The impact of baryons on the effective density contrast (5.9) as a function of the scale factor.}
\label{Fig5}
\end{figure}
\end{center}

In {}Fig.~\ref{Fig5} we show the evolution of the density contrast considering that both bulk and shear viscosities are present in the cosmic matter with dimensionless viscous parameters $\tilde{\xi} = 10^{-5}$ and $\tilde{\eta} = 10^{-5}$ (i.e., $R=1$). The dashed-dotted line corresponds to the case in which the entire matter is viscous (i.e., $\Delta_b=0$). The solid line corresponds to the case where baryons are accounted for, following Eq. (\ref{deltaeff}). In both cases we notice that the influence of the background expansion is equivalent. This occurs (as already mentioned) because viscosity values of order $\tilde{\xi}\sim 10^{-5}$ do not lead to a relevant deviation from the standard pressureless dark matter background scaling $\rho \sim a^{-3}$.
Nevertheless, we note that in the absence of standard pressureless baryons the growth suppression in $\Delta_{eff}$ is not relevant. Therefore, we can conclude that the inclusion of baryons tends to lead to slightly different upper
bounds on the dark matter viscosity. If the impact of baryons in the total matter clustering is subdominant we can also conclude that any property assigned to the baryonic sector e.g., viscosities or pressure, should not change the main conclusion of this work which has been based mainly on qualitative grounds.

\begin{center}
\begin{figure}[!htb]
\includegraphics[width=8cm]{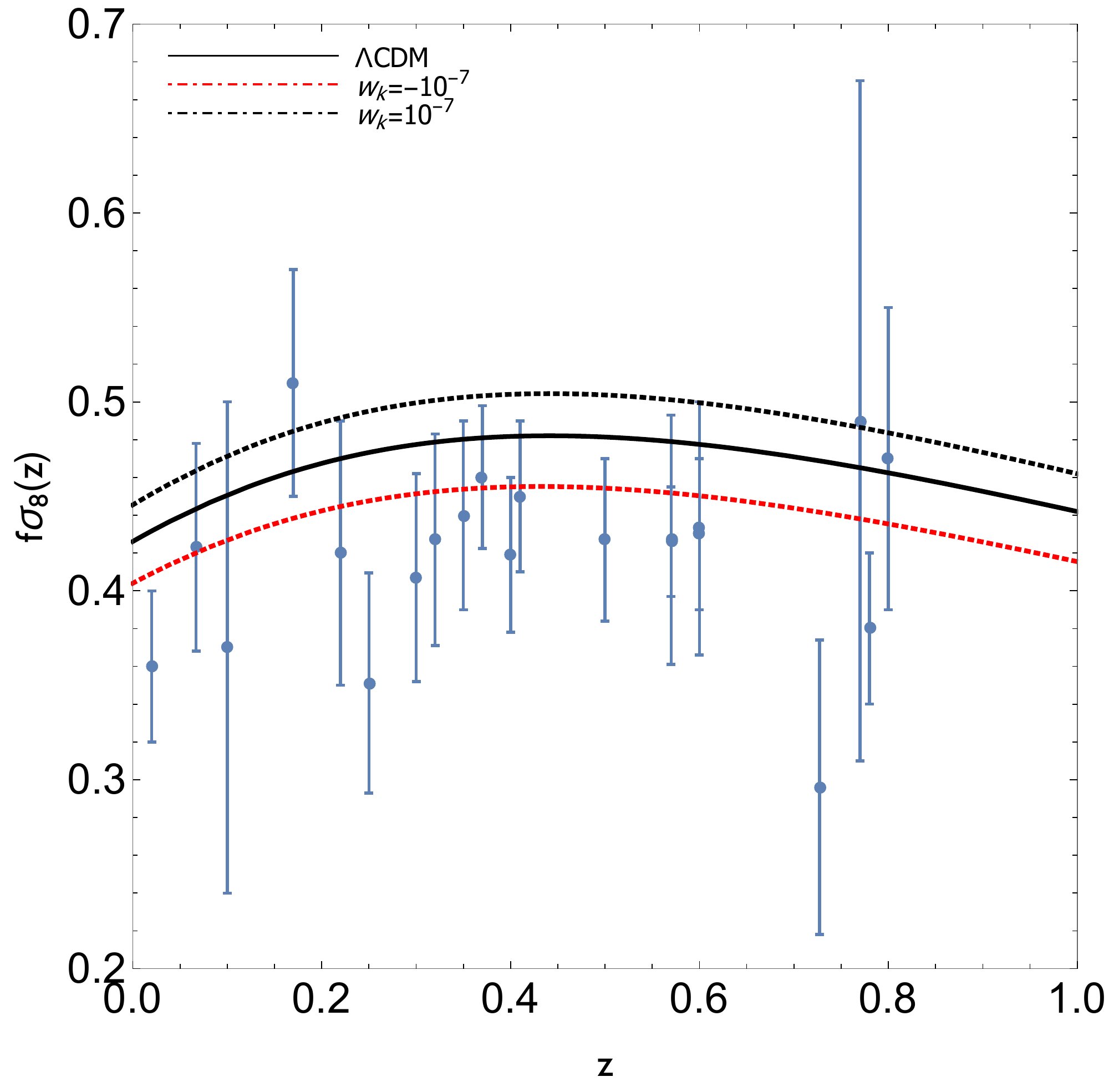}
\caption{Impact of the kinetic pressure.}
\label{Fig6}
\end{figure}
\end{center}

It is also worth mentioning that the matter fluid could possesses a tiny kinetic pressure. Indeed, the structure formation analysis constrains severely the magnitude of the parameter $w_k$ \cite{Xu:2013mqe}. The equations of the evolution of the matter contrast in the case are slightly different from Eqs.~(\ref{Deltab}) and (\ref{eulerv}) and are widely know in the literature~\cite{Xu:2013mqe}. {}For example, {}Fig.~\ref{Fig6} shows the impact of values $w_k= \pm 10^{-7}$.   
From the results shown in {}Fig.~\ref{Fig6}, we notice that the kinetic pressure leads to an uniform redshift independent displacement in the $f(z)\sigma_8 (z)$ evolution.
This effect should be contrasted with the one produced, e.g., by the shear viscosity
one shown in {}Fig~\ref{Fig3}(a), which acts in a more pronounced way at low redshifts.
Thus, the effect of the kinetic pressure can, in principle, be distinguished from that of the shear viscosity as more accurate cosmological data become available in the future. 

\section{Conclusions}
\label{conclusion}

We have studied in the present work the potential differences in the
Newtonian scalar potentials $\Psi$ and $\Phi$ as resulting from both a
possible deviation from the GR description for gravity and also by
considering that the anisotropic stresses in the energy momentum
tensor yield to $\Psi \neq \Phi$. The latter effect due to a shear
viscosity that dark matter might be endowed in the GR context.  {}For
this study, we have employed three different forms of parameterizing
the modified gravity effects through the modification of the Poisson
equation for the scalar potential $\Phi$. This is a strategy commonly
used in the literature to account the possible modifications generated
by different physical scenarios to GR. We have then contrasted these
modifications from modified gravity with those from the shear viscous
effects when added to GR. To gauge these modifications in the context
of the $\Lambda$CDM model, we have made use of the
redshift-space-distortion based $f(z)\sigma_8(z)$ data, which gives a
convenient probe of these different effects at the level of the
perturbations.

Our results show that, in general, modified gravity and shear
viscosity have opposing effects on the $f(z)\sigma_8(z)$ predicted by
the $\Lambda$CDM model. While modified gravity tends to  enhance the
gravitational coupling compared to the GR, thus leading to a stronger
matter agglomeration and a larger $f \sigma_8$ compared to
$\Lambda$CDM, the shear viscosity contribution to GR acts oppositely.
This, thus, leads to an interesting possibility of the shear viscosity
effects in GR maskering those effects from modified gravity. We have
seen that this tends to happen mostly effectively  at low redshifts in
all three cases of parameterizations of modified gravity that we have
considered. This compensation effect is, however, less effective at
high redshifts. This points out then for a possible best way for
differentiating these effects in future astrophysical searches and
probes using high redshift data. In this case, very accurate data on
the matter clustering via the $f(z)\sigma_8(z)$ measurements might
then be able to distinguish the effects studied here. 

\acknowledgments

We thank CNPq (Brazil), CAPES (Brazil) and FAPES (Brazil) for partial
financial support.  R.O.R is partially supported by research grants
from Conselho Nacional de Desenvolvimento Cient\'{\i}fico e
Tecnol\'ogico (CNPq), grant No. 302545/2017-4 and Funda\c{c}\~ao
Carlos Chagas Filho de Amparo \'a Pesquisa do Estado do Rio de Janeiro
(FAPERJ), grant No.  E - 26/202.892/2017.


\end{document}